# Unconventional quantum Hall effect and Berry's phase of $2\pi$ in bilayer graphene

K. S. Novoselov[1], E. McCann[2], S. V. Morozov[1,3], V. I. Falko[2], M. I. Katsnelson[4],

A. K. Geim[1], F. Schedin[1] & D. Jiang[1]

[1]Manchester Centre for Mesoscience and Nanotechnology, University of Manchester, Manchester, M13 9PL, UK

[2]Department of Physics, Lancaster University, Lancaster, LA1 4YB, UK

[3]Institute for Microelectronics Technology, 142432, Chernogolovka, Russia

[4]Institute for Molecules and Materials, Radboud University of Nijmegen, Toernooiveld 1, 6525 ED Nijmegen, The Netherlands

**There are known two distinct types of the integer quantum Hall effect. One is the conventional quantum Hall effect, characteristic of two-dimensional semiconductor systems [1,2], and the other is its relativistic counterpart recently observed in graphene, where charge carriers mimic Dirac fermions characterized by Berry's phase $\pi$, which results in a shifted positions of Hall plateaus [3-9]. Here we report a third type of the integer quantum Hall effect. Charge carriers in bilayer graphene have a parabolic energy spectrum but are chiral and exhibit Berry's phase $2\pi$ affecting their quantum dynamics. The Landau quantization of these fermions results in plateaus in Hall conductivity at standard integer positions but the last (zero-level) plateau is missing. The zero-level anomaly is accompanied by metallic conductivity in the limit of low concentrations and high magnetic fields, in stark contrast to the conventional, insulating behavior in this regime. The revealed chiral fermions have no known analogues and present an intriguing case for quantum-mechanical studies.**

Figure 1 provides a schematic overview of the quantum Hall effect (QHE) behavior observed in bilayer (2L) graphene by comparing it with the conventional integer QHE. In the standard theory, each filled single-degenerate Landau level (LL) contributes one conductance quantum $e^2/h$ towards the observable Hall conductivity. The conventional QHE is shown in Fig. 1a, where plateaus in Hall conductivity $\sigma_{xy}$ make up an uninterrupted ladder of equidistant steps. In 2L graphene, QHE plateaus follow the same ladder but the plateau at zero $\sigma_{xy}$ is markedly absent (Fig. 1b). Instead, the Hall conductivity undergoes a double-sized step across this region. In addition, longitudinal conductivity $\sigma_{xx}$ in 2L graphene remains of the order of $e^2/h$, even at zero $\sigma_{xy}$. The origin of the unconventional QHE behavior lies in the coupling between two graphene layers, which transforms massless Dirac fermions, characteristic of single-layer (1L) graphene [3-9], (Fig. 1c) into a novel type of chiral quasiparticles. Such quasiparticles have an ordinary parabolic spectrum $\varepsilon(p)=p^2/2m$ with effective mass $m$ but accumulate Berry's phase of $2\pi$ along cyclotron trajectories. The latter is shown to be related to a peculiar quantization where the two lowest LLs lie exactly at zero energy $\varepsilon$, leading to the missing plateau and double step shown in Fig. 1b.

Bilayer films studied in this work were made by the micromechanical cleavage of crystals of natural graphite which was followed by the selection of 2L flakes by using a combination of optical microscopy and atomic force microscopy as described in refs. [10,11]. Multi-terminal field-effect devices (inset in Fig. 2) were made from the selected flakes by using standard microfabrication techniques. As a substrate we used an oxidized heavily-doped Si wafer which allowed us to apply gate voltage $V_g$ between graphene and the substrate. The studied devices exhibited an ambipolar electric field effect such that electrons and holes could be induced in concentrations $n$ up to $10^{13}$cm$^{-2}$ ($n=\alpha \cdot V_g$ where $\alpha \approx 7.3 \cdot 10^{10}$cm$^{-2}$/V for a 300 nm SiO$_2$ layer). For further details about microfabrication of graphitic field-effect devices and their measurements we refer to the earlier work [3,4,10,11].

Figure 2a shows a typical QHE behavior in 2L graphene at a fixed gate voltage (fixed $n$) and varying magnetic field $B$ up to 30T. Pronounced plateaus are clearly seen in Hall resistivity $\rho_{xy}$ in high $B$, and they are accompanied by zero longitudinal resistivity $\rho_{xx}$. The observed sequence of the QHE plateaus is described by $\rho_{xy}=h/4Ne^2$, which is the same sequence as expected for a two-dimensional (2D) free-fermion system with double spin and double valley degeneracy [1,2,12-15]. However, a clear difference between the conventional and reported QHE emerges in the regime of small filling factors $\nu<1$ (see Figs. 2b,c and 3). This regime is convenient to study by fixing $B$ and varying concentrations of electrons and holes passing through the neutrality point $|n|\approx0$ where $\rho_{xy}$ changes its sign and, nominally, $\nu=0$. Also, because carrier mobilities $\mu$ in graphitic films are weakly dependent on $n$ [3,4,10], measurements in constant $B$ are more informative. They correspond to a nearly constant parameter $\mu B$, which defines the quality of Landau quantization, and this allows simultaneous observation of several QHE plateaus during a single voltage sweep in moderate magnetic fields (Fig. 2b). The periodicity $\Delta n$ of quantum oscillations in $\rho_{xx}$ as a function of $n$ is defined by the density of states $gB/\phi_0$ on each LL [1-10] (see Fig. 1). In Fig. 2c, for example, $\Delta n \approx 1.2 \cdot 10^{12} \text{cm}^{-2}$ at $B=12\text{T}$, which yields $g=4$ and confirms the double spin and double valley degeneracy expected from band structure calculations for 2L graphene [14,15].

Fig. 2b shows that, although the Hall plateaus in 2L graphene follow the integer sequence $\sigma_{xy}=\pm(4e^2/h)N$ for $N \geq 1$, there is no sign of the zero-$N$ plateau at $\sigma_{xy}=0$, which is expected for 2D free-fermion systems [1,2] (Fig. 1a). In this respect, the behavior resembles the QHE for massless Dirac fermions (Fig. 1c), where also there is no plateau but a step occurs when $\sigma_{xy}$ passes the neutrality point. However, in 2L graphene, this step has a double height and is accompanied by a central peak in $\rho_{xx}$, which is twice broader than all other peaks (Fig. 2c). The broader peak yields that in 2L graphene the transition between the lowest hole and electron Hall plateaus requires twice the amount of carriers needed for the transition between the other QHE plateaus. This implies that the lowest LL has double degeneracy $2 \times 4B/\phi_0$, which can be viewed as two LLs merged together at $n \approx 0$ (see LL charts in Fig. 1).

Continuous measurements through $\nu=0$ as shown in Fig. 2b,c have been impossible for conventional 2D systems where the zero-level plateau in $\sigma_{xy}=\rho_{xy}/(\rho_{xy}^2+\rho_{xx}^2)$ is inferred [1,2] from a rapid (often exponential) increase in $\rho_{xx}>>h/e^2$ with increasing $B$ and decreasing temperature $T$ for filling factors $\nu<1$, indicating an insulating state. To provide a direct comparison with the conventional QHE measurements, Fig. 3 shows $\rho_{xx}$ in 2L graphene as a function of $B$ and $T$ around zero $\nu$. 2L graphene exhibits little magnetoresistance or temperature dependence at the neutrality point, in striking contrast to the conventional QHE behavior. This implies that $\sigma_{xy}$ in 2L graphene does not vanish over any interval of ν and reaches zero only at one point, where $\rho_{xy}$ changes its sign. Note that $\rho_{xx}$ surprisingly maintains a peak value $\approx h/ge^2$ in fields up to 20T and temperatures down to 1K. A finite value of $\rho_{xx} \approx h/4e^2$ in the limit of low carrier concentrations and at zero $B$ was previously reported for 1L graphene [3]. This observation was in qualitative agreement with theory, which attributes the finite metallic conductivity and the absence of localization to the relativistic-like spectrum of 1L graphene (see refs. in [3]). 2L graphene has the usual parabolic spectrum, and the observation of the maximum resistivity $\approx h/4e^2$ and, moreover, its weak dependence on $B$ in this system is most unexpected.

The unconventional QHE in 2L graphene originates from peculiar properties of its charge carriers that are chiral fermions with a finite mass, as discussed below. First, we have calculated the quasiparticle spectrum in 2L graphene by using the standard nearest-neighbour approximation [12]. For quasiparticles near the corners of the Brillouin zone known as K-points, we find $\varepsilon(p) = \pm \frac{1}{2}\gamma_1 \pm \sqrt{\frac{1}{4}\gamma_1^2 + v_F^2 p^2}$, where $v_F = \frac{\sqrt{3}}{2}\gamma_0 a/\hbar$, $a$ is the lattice periodicity and $\gamma_0$ and $\gamma_1$ are the intra- and inter-layer coupling constants, respectively [13]. This dispersion relation (plotted in Fig. 2c) is in agreement with the first-principle band-structure calculations [14] and, at low energies, becomes parabolic $\varepsilon = \pm p^2/2m$ with $m=\gamma_1/2v_F^2$ (sign $\pm$ refers to electron and hole states). Further analysis [15] shows that quasiparticles in 2L graphene can be described by using the effective Hamiltonian

$$\hat{H}_2 = -\frac{1}{2m}\begin{pmatrix} 0 & (\hat{\pi}^+)^2 \\ \hat{\pi}^2 & 0 \end{pmatrix} \quad \text{where} \quad \hat{\pi} = \hat{p}_x + i\hat{p}_y. \tag{1}$$

$\hat{H}_2$ acts in the space of two-component Bloch functions (further referred to as pseudospins) describing the amplitude of electron waves on weakly-coupled nearest sites *A1* and *B2* belonging to two nonequivalent carbon sublattices *A* and *B* and two graphene layers marked as *1* and *2*.

For a given direction of quasiparticle momentum $\mathbf{p}=(p\cos\varphi, p\sin\varphi)$, Hamiltonian $\hat{H}_J$ of a general form $\begin{pmatrix} 0 & (\hat{\pi}^+)^J \\ \hat{\pi}^J & 0 \end{pmatrix}$ can be rewritten as

$$\hat{H}_J = \varepsilon(p)\vec{\sigma}\cdot\vec{n}(\varphi) \tag{2}$$

where $\vec{n} = -(\cos J\varphi, \sin J\varphi)$ and vector $\vec{\sigma}$ is made from Pauli matrices [15]. For 2L graphene, *J*=2, but notation *J* is useful because it also allows Eq. (2) to be linked with the case of 1L graphene, where *J*=1. The eigenstates of $\hat{H}_J$ correspond to pseudospins polarized parallel (electrons) or antiparallel (holes) to the 'quantization' axis $\vec{n}$. An adiabatic evolution of such pseudospin states, which accompanies the rotation of momentum $\vec{p}$ by angle $\varphi$, also corresponds to the rotation of axis $\vec{n}$ by angle $J\varphi$. As a result, if a quasiparticle encircles a closed contour in the momentum space (that is $\varphi = 2\pi$), a phase shift $\Phi=J\pi$ known as Berry's phase is gained by the quasiparticle's wavefunction [16]. Berry's phase can be viewed as arising due to rotation of pseudospin, when a quasiparticle repetitively moves between different carbon sublattices (*A* and *B* for 1L graphene, and *A1* and *B2* for 2L graphene).

For fermions completing cyclotron orbits, Berry's phase contributes to the semiclassical quantization and affects the phase of Shubnikov-de Haas oscillations (SdHO). For 1L graphene, this results in a $\pi$-shift in SdHO and a related ½-shift in the sequence of QHE plateaus [3-9], as compared to the conventional 2D systems where Berry's phase is zero. For 2L graphene, $\Phi=2\pi$ and there can be no changes in the quasiclassical limit ($N \gg 1$). One might also expect that phase $2\pi$ cannot influence the QHE sequencing. However, the exact analysis (see Supplementary Information) of the LL spectra for Hamiltonian $\hat{H}_J$ exhibiting Berry's phase $J\pi$ shows that there is an associated $J$-fold degeneracy of the zero-energy Landau level (that is Berry's phase of $2\pi$ leads to observable consequences in the quantum limit $N=0$). For the free-fermion QHE systems (no Berry's phase), $\varepsilon_N = \hbar\omega_c(N+1/2)$ and the lowest state lies at finite energy $\hbar\omega_c/2$, where $\omega_c = eB/m$. For 1L graphene ($J=1$; $\Phi=\pi$), $\varepsilon_N = \pm v_F\sqrt{2e\hbar BN}$ and there is a single state $\varepsilon_0$ at zero energy [5-9]. For 2L graphene ($J=2$; $\Phi=2\pi$), $\varepsilon_N = \pm\hbar\omega_c\sqrt{N(N-1)}$ and two states $\varepsilon_0=\varepsilon_1$ lie at zero energy [15].

The existence of a double-degenerate LL explains the unconventional QHE found in 2L graphene. This LL lies at the border between electron and hole gases and, taking into account the quadruple spin and valley degeneracy, it accommodates carrier density $8B/\phi_0$. With reference to Fig. 1, the existence of such LL implies that there must be a QHE step across the neutrality point, similarly to the case of 1L graphene [3-9]. Due to the double degeneracy, it takes twice the amount of carriers to fill it (as compared to all other LLs), so that the transition between the corresponding QHE plateaus must be twice wider (that is $8B/\phi_0$ as compared to $4B/\phi_0$). Also, the step between the plateaus must be twice higher, that is $8e^2/h$ as compared to $4e^2/h$ for the other steps at higher carrier densities. This is exactly the behavior observed experimentally.

In conclusion, 2L graphene adds a new member to the small family of QHE systems, and its QHE behavior reveals the existence of massive chiral fermions with Berry's phase $2\pi$, which are distinct from other known quasiparticles. The observation of a finite metallic conductivity $\approx e^2/h$ at filling factors $\nu \approx 0$ poses a serious challenge for theory.

FIGURE CAPTIONS

Figure 1. Three types of the integer quantum Hall effect. The drawings illustrate schematically the conventional integer QHE found in 2D semiconductor systems (**a**) (incorporated from refs. [1,2]) and the QHE in 2L graphene described in the present paper (**b**). Plateaus in Hall conductivity $\sigma_{xy}$ occur at values $(ge^2/h)N$ where $N$ is integer, $e^2/h$ the conductance quantum and $g$ the system degeneracy. The distance between steps along the concentration axis is defined by the density of states $gB/\phi_0$ on each LL, which is independent of a 2D spectrum [1-9]. Here, $B$ is magnetic field and $\phi_0=h/e$ the flux quantum. The corresponding sequences of Landau levels as a function of carrier concentrations $n$ are shown in blue and orange for electrons and holes, respectively. For completeness, (**c**) also shows the QHE behavior for massless Dirac fermions in 1L graphene.

Figure 2. QHE in bilayer graphene. (**a**) – $\rho_{xy}$ and $\rho_{xx}$ measured as a function of $B$ for fixed concentrations of electrons $n \approx 2.5 \cdot 10^{12}$ cm$^{-2}$ induced by the electric field effect. Inset: Scanning electron micrograph of one of more than ten bilayer devices studied in our work. The width of the Hall bar (dark central area) is $\approx 1\mu$m. (**b, c**) - $\sigma_{xy}$ and $\rho_{xx}$ are plotted as a function of $n$ at a fixed $B$ and temperature $T$=4K. Positive and negative $n$ correspond to field-induced electrons and holes, respectively. The Hall conductivity $\sigma_{xy}=\rho_{xy}/(\rho_{xy}^2+\rho_{xx}^2)$ was calculated directly from experimental curves for $\rho_{xy}$ and $\rho_{xx}$. $\sigma_{xy}$ allows one to see more clearly the underlying sequences of QHE plateaus. $\sigma_{xy}$ crosses zero without any sign of the zero-level plateau that would be expected for a conventional 2D system. The inset shows the calculated energy spectrum for 2L graphene, which is parabolic at low $\varepsilon$. Carrier mobilities $\mu$ in our 2L devices were typically $\approx$3,000 cm$^2$/Vs, which is lower than for devices made from 1L graphene [3,4]. This is surprising because one generally

expects more damage and exposure in the case of 1L graphene that is unprotected from the immediate environment from both sides.

Figure 3. Resistivity of 2L graphene near zero concentrations as a function of magnetic field and temperature. The peak in $\rho_{xx}$ remains of the order of $h/4e^2$, independent of $B$ (**a,b**) and $T$ (**c,d**). This yields no gap in the Landau spectrum at zero energy. (**b**) - For a fixed $n \approx 0$ and varying $B$, we observed only small magnetoresistance. The latter varied for different devices and contact configurations (probably indicating the edge state transport) and could be non-monotonic and of random sign. However, the observed magnetoresistance never exceeded a factor of 2 in any of our experiments for undoped 2L graphene.

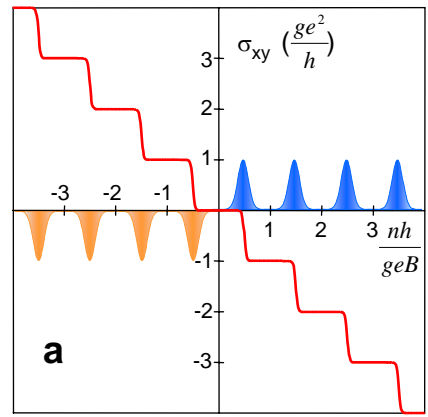
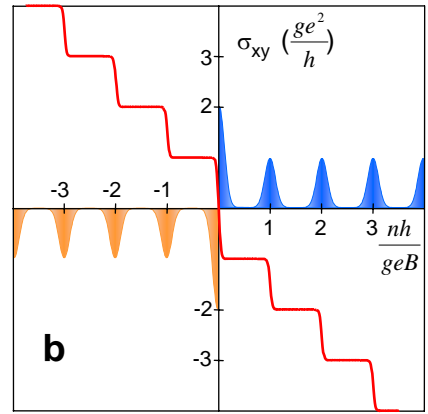
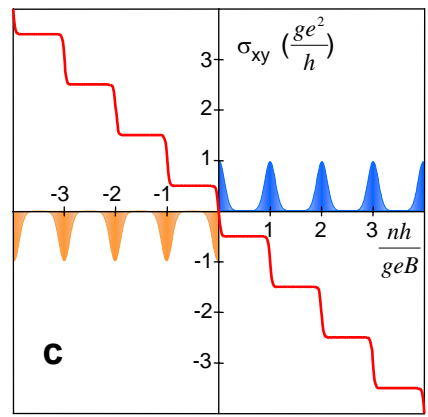

Figure 1

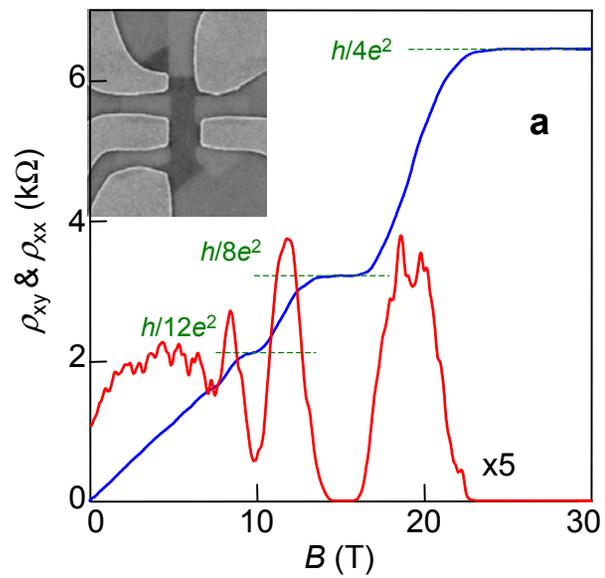
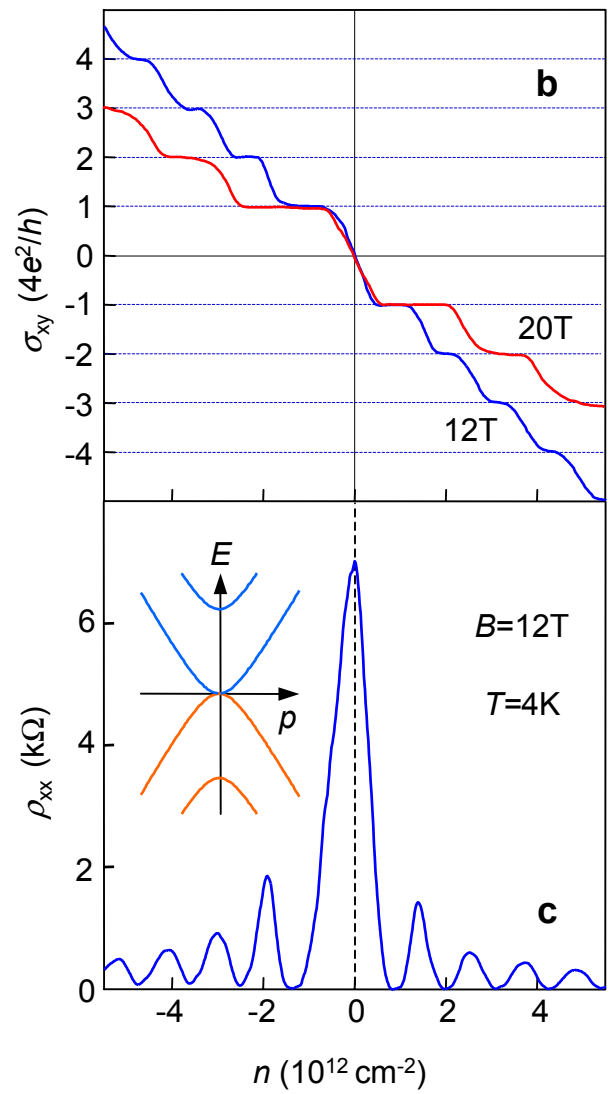

Figure 2

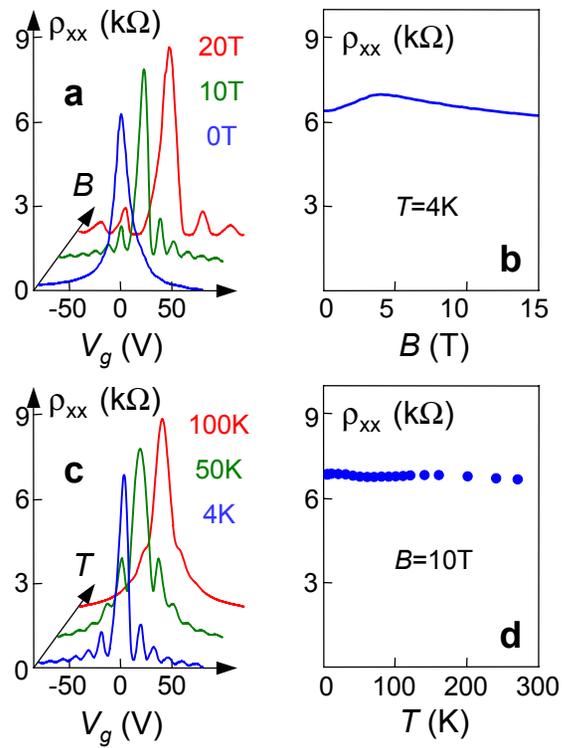

Figure 3